# Enhanced Granular Magnetoresistance due to Ferromagnetic Layers

J. Balogh[(1)], M. Csontos[(2)], D. Kaptás[(1)] and G. Mihály[(2)]

[1]*Research Institute for Solid State Physics and Optics, 1525 Budapest P.O.B. 49., Hungary*
[2]*Department of Physics, Budapest University of Technology and Economics,
1111 Budapest, Budafoki út 8, Hungary*

Giant magnetoresistance (GMR) of sequentially evaporated Fe-Ag structures have been investigated. Direct experimental evidence is given that inserting ferromagnetic layers into a granular structure significantly enhances the magnetoresistance. The increase of the GMR effect is attributed to spin polarization effects. The large enhancement (up to more than a fourfold value) and the linear variation of the GMR in low magnetic fields are explained by scattering of the spin polarized conduction electrons on paramagnetic grains.

Giant magnetoresistance (GMR) in granular composites[1, 2] is explained mainly by the elastic spin dependent scattering of conduction electrons at the interface between magnetic and nonmagnetic regions. The applied field changes the magnetic configuration of the system, which also depends on the temperature and the grain size distribution of the magnetic particles. At high enough temperatures all the particles are superparamagnetic and the magnetoresistance is proportional to the square of the magnetization of the sample[2]. When decreasing the temperature below the blocking temperature a certain fraction determined by the grain size distribution can be aligned ferromagnetically with the applied field. The presence of ferromagnetically aligned particles was supposed to result in a linear term[3] when the magnetoresistance is expressed as a function of the macroscopic magnetization of the sample. However, the possibility that the inherent magnetic heterogeneity of such granular samples may have an influence on the magnitude of the effect was raised only recently[4]. It was shown theoretically[4], that the spin-dependent scattering on paramagnetic impurities and/or small paramagnetic clusters in the nonmagnetic metal is strongly enhanced in heterogeneous magnetic systems as compared to nonmagnetic metals with dissolved magnetic impurities. The enhancement is due to the difference in the resistance of the spin-up and spin-down conduction channel, which does not cancel out if the spin diffusion length is sufficiently large. The increase of the GMR effect observed at low temperatures in granular Co-Ag films[4] was explained by this mechanism. In this letter we present a direct experimental evidence of this sort of enhancement and show that inserting ferromagnetic Fe layers into Fe-Ag granular films can significantly increase the room temperature magnetoresistance.

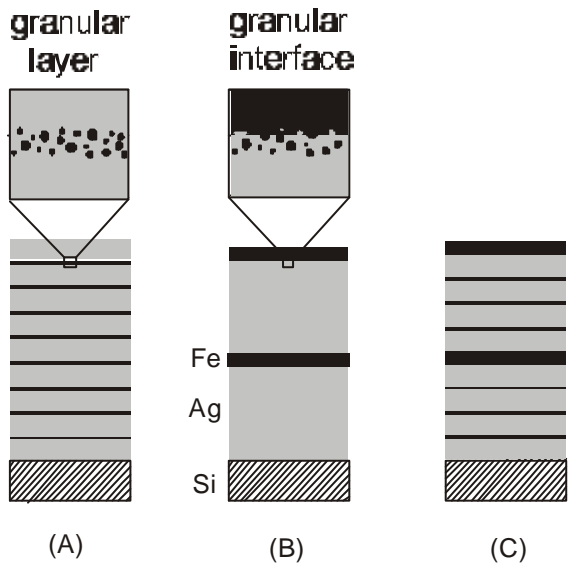

*Fig. 1 Illustration of the basic structural properties of the multilayer samples labeled as A, B and C.*

The samples examined are Fe/Ag multilayers vacuum evaporated in a base pressure of $10^{-7}$ Pa onto Si single crystal substrate at room temperature with the following sequences:

  A)   [Ag (2.6 nm) / Fe (0.2 nm)]$_{75}$ / Ag (2.6 nm)
  B)   [Ag (10.4 nm) / Fe (1.5 nm)]$_{32}$ / Ag (10.4 nm)
  C)   [[Ag (2.6 nm) / Fe (0.2 nm)]$_3$ / [Ag (2.6 nm) / Fe (1.5 nm)]]$_{32}$ / Ag (2.6 nm)

The basic units of the above three multilayers are shown in Fig. 1.

The magnetic structure of the samples was examined by transmission Mössbauer spectroscopy. For this aim the evaporated films had been removed from the substrate and folded up to achieve an appropriate thickness. The Mössbauer spectra measured at room temperature are shown in Fig. 2. The 0.2 nm Fe layer in sample A) is not continuous and the sample is superparamagnetic[5] with a blocking temperature around 50 K. The room temperature Mössbauer spectrum (Fig. 2a) consists of a paramagnetic doublet with a large isomer shift relative to α-Fe (0.18 mm/s) and a quadruple splitting (0.45 mm/s) characteristic to small Fe clusters in an Ag matrix. The granular structure is illustrated in the magnified box of Fig. 1A.

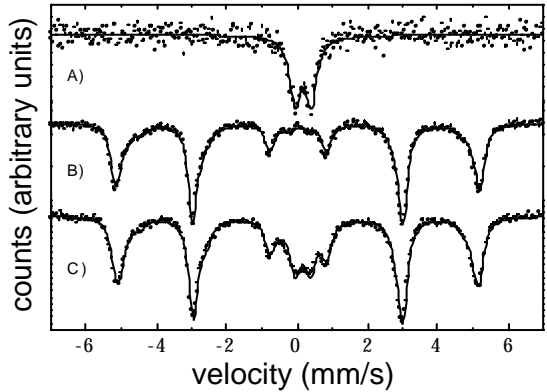

Fig. 2 Room temperature Mössbauer spectra of samples A, B, and C.

The 1.5 nm Fe thickness of sample B) is large enough to form a continuous ferromagnetic layer, as it was indicated by X-ray reflectivity and magnetization measurements[6]. The Mössbauer spectrum (Fig. 2b) can be fitted by magnetically split components[5] belonging to different Fe positions at the Fe/Ag interface, in good agreement with other results[7]. Note that the magnetic splitting of the main spectral component (32.5 Tesla) is slightly but significantly lower than the bulk Fe value (33 Tesla), which may be attributed to a reduction of the Curie temperature[5] due to Ag impurities in the Fe layers. On the other hand the presence of a granular interface, i.e. Fe impurity atoms and clusters in the Ag layer, was shown to exist even in case of a 25 nm thick Fe layer[6]. The granular interface of a 1.5 nm thick Fe layer is illustrated in Fig. 1B.

Sample C) contains both granular and continuous Fe layers in a superposition of the periodicity of sample A) and B). The Mössbauer spectrum (Fig. 2c) of this sample is a combination of the spectra measured on samples A) and B) and the spectral fraction of the paramagnetic component (22 %) is in satisfactory agreement with that expected according to the nominal sample structure (28.6 %). These results show that alteration of the overall sample structure does not effect significantly the individual layer properties.

Magnetoresistance of the samples measured at room temperature by four contact method in transversal magnetic field geometry is shown in Fig. 3. The magnetoresistance of sample A) with granular Fe layers is very similar to that observed[8] on co-evaporated Fe-Ag granular films. The magnetoresistance measured in parallel and transversal magnetic fields are equally large[6] and does not saturate up to 12 T applied magnetic field. The shape of the curve can be basically described[6] by the model of Gittleman[9] generally accepted to explain the magnetoresistance of granular materials.

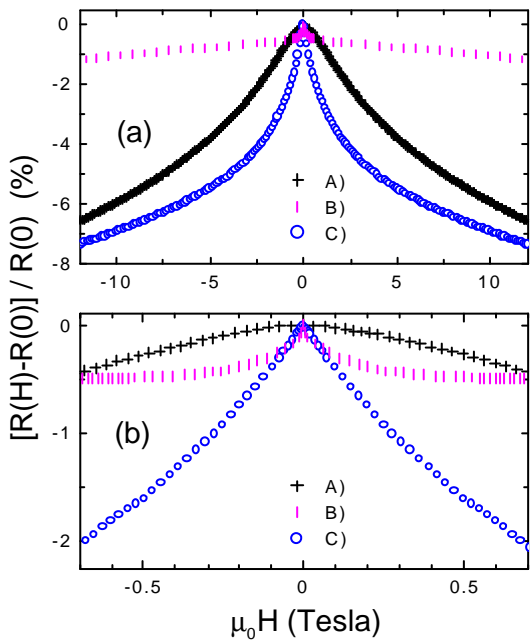

Fig. 3 Magnetoresistance of samples A, B, and C measured at room temperature (a), and the low field range on a magnified scale (b)

The magnetoresistance of sample B) with continuous Fe layers is much smaller but apart from a small anisotropic contribution below 0.1 T (not shown) it exhibits similar behavior. This field dependence has been attributed[6] to the granular interface. Note that the thickness of the Ag spacer in sample B) is large enough that exchange coupling is negligible[10] between the Fe layers.

The magnetoresistance measured on sample C) with the superimposed paramagnetic and ferromagnetic layers is considerably enhanced as compared either to sample A) or B) in almost the whole investigated magnetic field range, as it is shown in Fig. 3a. The shape of the field dependence is also modified by the ferromagnetic layers. The most remarkable change is observed at low fields, as it is shown in Fig. 3b. While the magnetoresistance of sample A) exhibits a smooth peak at zero field, in case of sample C) it starts almost linearly up to about 0.5 T. Moreover, at 0.7 T it increases by a factor of 4.7 and 4.2 as compared to the value in sample A) and B), respectively. Presumably, the enhancement can be further increased by choosing an optimal thickness and periodicity of the ferromagnetic layers, however, this needs more detailed investigations.

The electron scattering on randomly distributed small magnetic clusters in a nonmagnetic matrix leads to a resistance contribution[11]:

$$r_{\uparrow\downarrow} = r(1 \pm \alpha J + \beta J^2) \qquad (1)$$

where $\rho$ is the resistance coming from the spin independent scattering, J is the s-d exchange interaction, and $\alpha$ and $\beta$ are field dependent prefactors given in Ref(10). Note that the sign of the linear term depends on the spin direction of the electron. For the paramagnetic sample the linear term is averaged out, and the magnetoresistance contains only the term second order in J. This is the situation in case of sample A), where the magnetoresistance is attributed to scattering on small paramagnetic Fe clusters.

The presence of ferromagnetic layers destroys the spin symmetry in sample C). The conduction electrons become spin polarized giving rise to the appearance of the first order term of Eq(1) in the macroscopic resistance. This modifies both the shape and the magnitude of the magnetoresistance. A phenomenological description of the enhancement can be given by introducing spin dependent resistance terms arising from the scattering on the oriented Fe layers ($R_\uparrow$ and $R_\downarrow$). If the spin diffusion length is larger than the electron mean free path and the distance between the magnetic layers, the resistance of the spin up and spin down channel is to be calculated separately as the series of the layer and grain resistances for each channel; $R_\uparrow+r_\uparrow$ and $R_\downarrow+r_\downarrow$. In the above two current model[12] the total resistance of the sample is:

$$R = \frac{(R_\uparrow + r_\uparrow)(R_\downarrow + r_\downarrow)}{(R_\uparrow + r_\uparrow) + (R_\downarrow + r_\downarrow)} = \frac{R_\uparrow - R_\downarrow}{R_\uparrow + R_\downarrow} r\alpha J + \frac{R_\uparrow^2 + R_\downarrow^2}{(R_\uparrow + R_\downarrow)^2} r(1 + \beta J^2) . \qquad (2)$$

The expansion in powers of r in Eq(2) is expected to be valid in the low field regime as $R_{\uparrow\downarrow}$ is field independent, while $r_{\uparrow\downarrow} \to 0$ as $H \to 0$. (We assume that the magnetic moments of the Fe layers are fully aligned in the field range investigated.) The field dependence of the first order term in J is determined by the magnetization of the grains. For $kT \gg g\mu_B H$ it is linear in H, being proportional to $S(S+1)g\mu_B H/kT$ (see Ref. 1). The non-vanishing first order term in J explains both the observed enhancement and the remarkable change in the shape of the field dependence when ferromagnetic layers are added to the granular structure.

In conclusion we have shown that the magnetoresistance of Fe-Ag granular layers can be enhanced by inserting ferromagnetic layers with appropriate spacing. The largest enhancement, up to 4.7 times the value in the granular reference sample, is observed at low fields. The effect is explained by spin polarization of the conduction band by the ferromagnetic layers giving rise to a scattering on the paramagnetic grains which is first order in J.

**Acknowledgements**
This work was supported by the Hungarian Research Funds OTKA T026327, OTKA T034602, and NKFP 3-064-2001.